\documentclass[aps,prapplied,twocolumn,superscriptaddress]{revtex4-2}

\usepackage{bm} 		
\usepackage{amsmath, graphicx} 
\usepackage{xcolor,soul}

\begin{document} 

\title{Connection between antennas, beam steering, and the moir\'{e} effect} 
\author{B. H. McGuyer}
\affiliation{Amazon.com, Inc., 410 Terry Ave. North, Seattle, WA 98109.}
\author{Qi Tang}
\affiliation{Google Inc., 1600 Amphitheatre Parkway, Mountain View, CA 94043. }
\date{\today}
\begin{abstract}
The moir\'{e} effect provides an interpretation for the steering of antennas that form beams through internal spatial interferences. 
We make an explicit connection between such antennas and the moir\'{e} effect, and use it to model six planar antennas that steer by scaling, rotating, or translating operations. 
Three of the antennas illustrate 
how to use moir\'{e} patterns to generate antenna designs. 
\end{abstract}
\maketitle


\section{Introduction} 
Electromagnetic antennas are a vital and ubiquitous technology in modern communication systems and a multitude of other applications. 
With a long history, research and development continues to create novel antennas and to improve their designs. 
The challenges of next-generation communication systems still drive much of this work. 
For example, systems that plan to use continuously moving satellites in constellations or high-altitude platforms will need sophisticated ground-based user antennas to continuously track and switch between moving terminals. 
Developing low-cost and low-power user antennas for those systems is one of many areas of current interest.

\begin{figure}[b!]
\centering
\includegraphics[width=0.94\columnwidth]{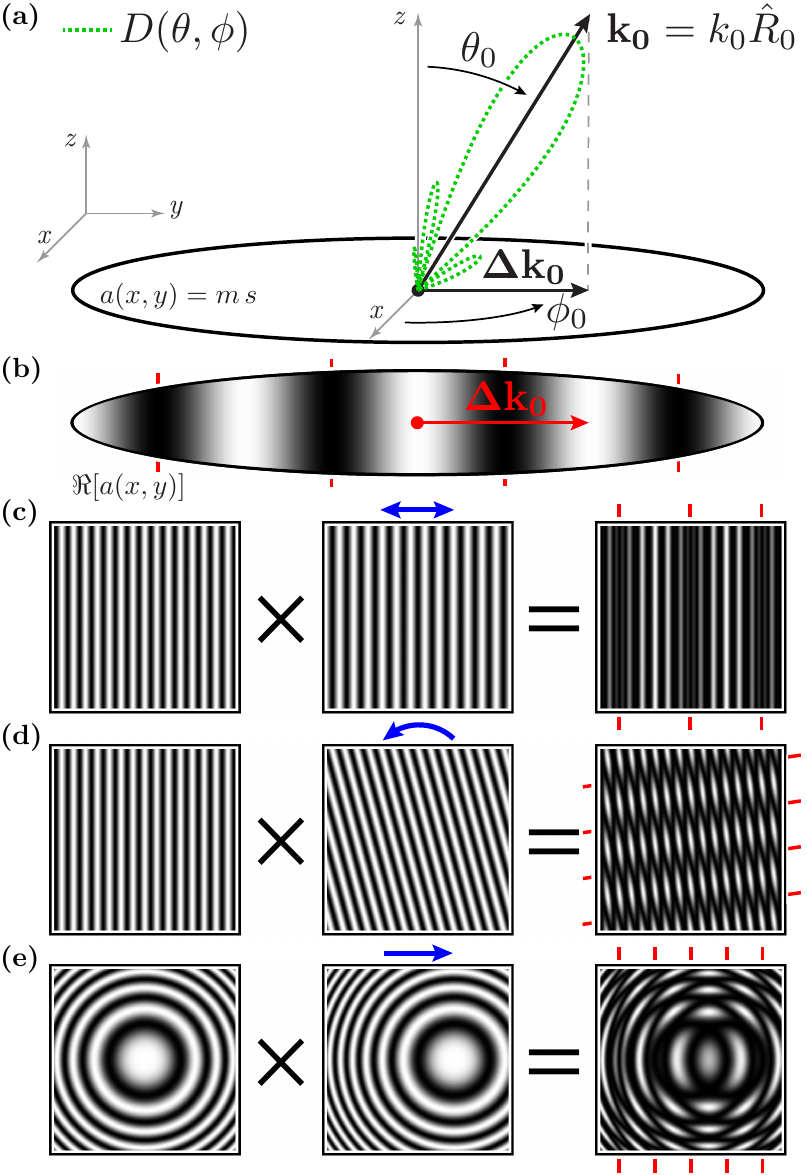}
\caption{ 	\label{fig1} 
Beam steering and moir\'{e} patterns. 
(a) Example circular antenna aperture showing conventions for beam angles and wavevectors. 
Includes a polar cross section of an example directivity in dB with a main lobe. 
(b) For ideal beam steering, a holographic projection of an aperture distribution shows tilt fringes given by an in-plane wavevector. 
Red lines highlight fringe nodes. 
Similar fringes are produced as moir\'{e} patterns from multiplying two identical images after one image is (c) scaled, (d) rotated, or (e) translated (blue arrows). 
}
\label{overview}
\end{figure}

There is a connection between antennas and the moir\'{e} effect that is known in optics \cite{edmund1,edmund2,2021rsi} but does not seem to be as appreciated or utilized in radio, microwave, or mm-wave applications. 
This connection exists if spatial interferences play a key role in the internal functions of an antenna, for example, to generate and dynamically control beams. 
The moir\'{e} effect is a general name for the production of such spatial interference patterns on surfaces when structures like two-dimensional images are combined \cite{amidror1,amidror2}. 
Well-known moir\'{e} patterns include the large-scale interferences that occur when nearly identical repetitive structures are superposed, such as the parallel lines or curvilinear rulings shown in Fig.~\ref{fig1}. 
Additional examples include aliasing, Glass patterns from aperiodic or random structures, and small-scale interferences like rosettes and dot trajectories. 
The moir\'{e} effect has many applications, from strain measurement to superresolution microscopy, 
in part because it can magnify small differences or repetitive structures \cite{amidror1,amidror2,sr1,patorski,walker,cloud}. 
Recently, it has been used to control the properties of bilayer graphene \cite{graphene} and of metamaterials \cite{moire-metasurface1,moire-metasurface2}.

In this paper, we make the connection between antennas and the moir\'{e} effect explicit and, as an example, illustrate how beam steering can be generated and controlled by moir\'{e} patterns. 
We begin with background on the relationship between antennas, Fourier analysis, and beam steering 
and adapt this background to discretized antennas (i.e., slot and phased arrays). 
We then illustrate how the dynamic beam steering of four antennas can be modeled as moir\'{e} patterns, two of which are potential candidates for next-generation satellite constellation user terminals \cite{VICTS1,stingarray}. 
One of these candidates is our recent experimental work \cite{stingarray} towards a low-cost and low-power phased-array antenna. 
Finally, we examine two additional moir\'{e} patterns 
relevant to beam steering, and conclude with a discussion of potential areas for future investigation.

\section{Antenna directivity} 

The directivity $D(\theta,\phi)$ of an antenna characterizes its ability to transmit and receive radiation in the far field in different directions, as sketched in Fig.~\ref{fig1}. 
It is a radiant intensity equal to the antenna power gain for ideal efficiency and its angular shape gives the radiation pattern of the antenna. 

For a finite planar antenna that is larger than the radiative wavelength $\lambda_0 = 2 \pi / k_0$ of interest, 
the directivity in the forward direction is very nearly given by the 
angular spectrum of a suitable  
field or current distribution $a(x,y)$ 
across a chosen antenna aperture 
that describes how the antenna produces or receives far-field radiation, 
\begin{align} 
D(\theta,\phi) 	\label{D}
	&\approx C_0 \, \left| K(\theta) \, \mathcal{F}[a](u,v) \right|^2  
\end{align} 
\cite{orfanidis,clarke,jull,bracewell,mailloux}. 
Here, the coefficient $C_0$ is set by normalization, usually so that 
$\int_0^{2 \pi} d\phi \int_0^\pi d\theta \sin(\theta) \, D(\theta,\phi) = 1$ 
when radiation in all directions and polarizations is included. 
The obliquity factor $K(\theta)$ varies slightly between theories, but is typically 
$\cos(\theta)$ or $[1 + \cos(\theta)]/2$ \cite{goodman,bracewell,jull,clarke}. 

For antennas producing narrow beams, the directivity (\ref{D}) is almost entirely due to the Fourier transform $\mathcal{F}[a](u,v)$ of the aperture distribution $a(x,y)$. 
In this work, the two-dimensional transform 
convention is 
\begin{align} 	\label{FT}
\mathcal{F}[a](u,v) 
	&=  \int_{-\infty}^\infty \int_{-\infty}^\infty a(x,y) \, e^{- i (x u + y v)} dx dy 
\end{align}
with spatial-frequency variables $u$ and $v$ that are wavenumbers in $k$-space. 
For radiation with a given wavenumber $k_0$, 
these variables give the in-plane directional cosines of the far-field direction unit vector as 
\begin{align} 	\label{R}
\hat{R}(\theta,\phi) 
	&= (u/k_0) \, \hat{x} + (v/k_0) \, \hat{y} + \cos(\theta) \, \hat{z} 
\end{align} 
and correspond to elevation and azimuth angles as 
\begin{align} 
u = k_0 \cos(\phi) \sin(\theta)  	\quad \text{and} \quad 
v = k_0 \sin(\phi) \sin(\theta) 	\label{uv} 
\end{align} 
with the conventions shown in Fig.~\ref{fig1}. 
For radiation with a particular direction $\hat{R}_0 = \hat{R}(\theta_0,\phi_0)$ set by the values $(u_0, v_0)$, or equivalently, the angles $(\theta_0, \phi_0)$, such as the peak of a main lobe, 
the free-space  
wavevector for that direction is 
${\bf k_0} = k_0 \hat{R}_0.$
In this convention, 
the inverse Fourier transform is 
$a(x,y) = (2 \pi)^{-2} \int_{-\infty}^\infty \int_{-\infty}^\infty \mathcal{F}[a](u,v) \, e^{i (u x + v y)} du dv.$

The aperture distribution $a(x,y)$ 
is a complex-valued, dimensionless function that characterizes the amplitude of oscillatory tangential fields or currents in the aperture plane. 
Ideally, it is zero outside the aperture, though in practice it may need to include significant fields outside the aperture. 
For a transmitting antenna, for example, that synthesizes an electric field ${\bf E}$ with in-plane polarization 
$\hat{\boldsymbol{\epsilon}}$ and angular frequency $\omega$, this distribution could give the field as 
${\bf E}(x,y,t) = \Re [ a(x,y) \, \hat{\boldsymbol{\epsilon}} \, e^{- i \omega t}]$. 
If an additional aperture distribution is needed to include an orthogonal polarization, the spectra of both distributions 
typically combine as intensities in the directivity. 
Note that additional techniques are generally needed to capture aperture boundary effects, which tend to be significant for small apertures \cite{jull,orfanidis}. 

The transform ${F}[a](u,v)$ is an angular spectrum of plane-wave radiation \cite{jull,clarke,bracewell,booker:1950}. 
For a free-space wavenumber $k_0$, the values that contribute to the directivity are located within a circle of radius $k_0$ about the origin known as the visibility window. 
Inside this window, the value at each point $(u,v)$ gives the complex amplitude of a plane wave with direction set by (\ref{R}) and (\ref{uv}). 
Outside this window, the values correspond to evanescent (inhomogeneous plane) waves that store reactive power but do not radiate. 

\subsection{Discretized antennas} 

Aperture distributions are most often used to model antennas with continuous apertures, for example, horn antennas with open apertures or reflector antennas with projected apertures \cite{clarke}. 
However, 
they also apply to discretized antennas such as phased arrays or slot arrays. 
In this work, we model such antennas as follows. 
For an antenna made of a repetitive pattern of identical elements, we approximate the aperture distribution by a two-variable convolution 
\begin{align}	\label{abc}
a(x,y) &\approx (b * c)(x,y)  \nonumber \\ 
	&= \int_{-\infty}^\infty \int_{-\infty}^\infty b(p,q) c(x-p,y-q) dp dq 
\end{align} 
of a distribution $b(x,y)$ for the elements and 
an array distribution $c(x,y)$ describing both where the elements are located and how their amplitudes and phases are controlled. 
This composition assumes the elements radiate independently. 
By construction, $c(x,y)$ includes all details needed for the element distributions to be identical, such as fixed phase offsets, in-plane element rotations or reflections, and amplitude control (e.g., tapering). 
For an antenna with multiple types of elements, the aperture distribution is a sum of convolutions. 

Using this convolution separates the directivity (\ref{D}) for a discretized antenna into a product 
\begin{align} 
D(\theta,\phi) 
	&\approx C_0 \, D_b(\theta,\phi) \, | AF(\theta,\phi)|^2 
\end{align} 
of an element directivity 
$D_b(\theta,\phi) \propto |K(\theta) \,\mathcal{F}[b](\theta,\phi)|^2$ 
and the squared magnitude of an array factor 
$AF(\theta,\phi) = \mathcal{F}[c](\theta,\phi)^*,$ 
where a star denotes complex conjugation \cite{mailloux}. 
For example, 
the array distribution and array factor for a phased array of $N$ elements with positions $(x_n, y_n)$ and complex-valued control amplitudes $c_n$ are 
$c(x,y) = \sum_{n=1}^N c_n \, \delta(x-x_n) \delta(y-y_n),$ 
where $\delta(x)$ is a Dirac delta function, and 
$AF(\theta,\phi) 
	= \sum_{n=1}^N c_n \, \exp[{i k (x_n u + y_n v)}].$ 
This follows the standard treatment of diffraction from gratings or aperture arrays in optics, and is similar to that of 
structure factors 
in crystallography. 
Note that the element directivity could describe elements that are not small compared to a wavelength. 

A discretized antenna that produces a narrow beam can be approximated as a continuous antenna with $a(x,y) \approx c(x,y)$ as follows. 
Broadly, 
if the elements provide little directivity themselves, 
such that $D_b(\theta,\phi)$ varies little within the region of interest, 
then the directivity is almost entirely due to $\mathcal{F}[c](\theta,\phi)$, or equivalently, the array factor. 
If the discretization is sufficiently dense, then the array distribution $c(x,y)$ contains the same information as the aperture distribution $a(x,y)$ within the bandlimit of the visibility window. 
For example, this holds for a rectangular array if the element spacings satisfy the Whittaker--Nyquist--Kotelnikov--Shannon sampling theorem. 
In this case, the aperture and array distributions are interchangeable in defining the antenna directivity. 
More narrowly, if the element directivities are not critical to the particular details of interest, namely main-lobe steering in this work, 
then this approximation holds only for modeling those details. 

For convenience, therefore, we will model both continuous and discretized antennas using $a(x,y)$ in the examples to follow. 
For the discretized antennas, however, it is to be understood that we approximate $a(x,y) \approx c(x,y)$ and consider only the element positions and their amplitude and phase control for modeling directivity and steering. 
This approximation applies because each of the examples satisfies one or both of the cases outlined above, and could be removed using (\ref{abc}).

\section{Connection with moir\'{e} effect} 
The aperture distribution $a(x,y)$ of a continuous or discrete antenna can be interpreted as a two-dimensional image with complex values. 
Moir\'{e} patterns often arise in images if they are a nonlinear combination of two or more images. 
Therefore, a connection exists with the moir\'{e} effect if the aperture distribution is produced by a nonlinear combination of two or more distributions. 
This connection is particularly important if the combination produces spatial interferences (or moir\'{e} patterns) that play a key role in the antenna, such as to synthesize and control dynamic beam steering. 

The traditional case is multiplying two input images to generate an output image \cite{amidror1,amidror2}. 
Fig.~\ref{fig1} shows three examples with greyscale images made from distributions with real values between zero and unity. 
For each, the two input images have sinusoidal linear or curvilinear rulings as distributions. 
Each output image has a distribution that is the point-by-point multiplication of the input distributions. 
The output images display moir\'{e} patterns, the most visible of which are large-scale interferences with nodes highlighted by red lines. 

Following the traditional case, we will consider aperture distributions that are produced by the multiplication 
\begin{align} 	\label{asm}
a(x,y) = s(x,y) \, m(x,y) 
\end{align} 
of a ``source'' distribution $s(x,y)$ and a ``mask'' distribution $m(x,y)$. 
This multiplication is nonlinear and generates spatial interferences in the angular spectrum $\mathcal{F}[a](u,v)$, which is given by a two-variable convolution of the mask and source spectra as 
\begin{align} 	\label{convasm}
\mathcal{F}[a](u,v) = \left( \mathcal{F}[s] * \mathcal{F}[m] \right)(u,v) / (2 \pi)^2. 
\end{align} 
Note that the interferences can be dynamically controlled if one or both of the input distributions or their means of nonlinear combination are parametrically modified, for example, in a reconfigurable antenna \cite{reconfigurable}. 

While many nonlinear combinations are possible, the traditional construction (\ref{asm}) approximately describes a wide range of antennas. 
We will illustrate this in examples to follow, but generally: 
For slot and phased arrays, the mask could describe the layout of the elements with their fixed phase and amplitude control and the source could describe their illumination. 
For transmitarrays and reflectarrays \cite{reflectarray,transmitarray}, the mask could describe a phase-shifting surface and the source could describe its illumination. 
For metasurface antennas \cite{metamaterial,metaRef2,metaRef3}, the mask could describe an artificially structured effective radiating surface and the source could describe an exciting surface wave or illumination.

The traditional construction (\ref{asm}), however, assumes the source and mask are independent and generally does not apply if the source and mask interact, unless their interaction can be included at least approximately (e.g., attenuation of transverse illumination as described below). 
For example, it does not apply to most cavity-backed antennas relying on multiple internal reflections.

The connection with the moir\'{e} effect relates primarily to the means of producing and controlling an aperture distribution. 
The same distribution $a(x,y)$ in (\ref{asm}) could be produced by other means. 
For example, a phased array could produce an equivalent distribution using phase shifters for each element. 
However, the connection presents an opportunity to explore or compare different possible implementations of an antenna. 
An electrically steered phased array, for example, typically uses a large number (e.g., hundreds) of phase-control degrees of freedom to steer the two angles of a beam direction. 
Examples to follow use the connection to explain how some planar antennas achieve similar control with only two mechanical degrees of freedom, like rotations or translations, potentially reducing the antenna power dissipation and cost. 
This opportunity for cost and power reduction was the motivation for our recent work \cite{stingarray}, which provides one of the examples to follow.

\section{Beam steering with moir\'{e} effect} 
Consider an antenna aperture that transmits a single narrow main-lobe beam that can be steered through different directions, such as sketched in Fig.~\ref{fig1}. 
In practice, the ideal aperture distribution 
that produces the highest directivity 
for a given direction $\hat{R}_0$ 
has a uniform amplitude $a_0$ but a linear phase gradient with a tip and tilt matching the intercepted phase of a plane wave along that direction, 
\begin{align} 		\label{a1} 
a_1(x,y) = a_0 \, \exp[{i (u_0 x + v_0 y)}] \, w(x,y), 
\end{align} 
where the aperture-stop window function $w(x,y)$ is unity inside the aperture and zero outside \cite{mailloux}. 
While more complicated distributions can produce narrower beams, such superdirectivity is generally challenging to implement \cite{mailloux}. 

The directivity (\ref{D}) follows from the spectrum $\mathcal{F}[a_1](u,v) = a_0 \, \mathcal{F}\left[w\right](u - u_0,v-v_0) $, which has an impulse located at $(u_0,v_0)$ with a shape given by the aperture-stop spectrum. 
Together with (\ref{R}) and (\ref{uv}), the impulse location gives the main-lobe beam angles as 
\begin{align}
\theta_0 	&= 
	\text{sgn}(u_0) \arcsin \sqrt{(u_0^2 + v_0^2 )/ k_0^2} 		
	\in (-\pi/2, \pi/2) 		\label{theta0} \\ 
\phi_0 	&= 
	\arctan (v_0 / u_0 )		 
	\in ( -\pi/2, \pi/2) 	\label{phi0}
\end{align} 
for a convention allowing negative elevation angles to avoid a discontinuity about broadside ($\theta_0 = 0$, or zenith). 
Converting to other conventions follows from 
$D(-\theta,\phi) = D(\theta,\phi + \pi)$ and 
$D(\theta,\phi) = D(\theta,\phi + 2\pi n)$ for integer $n$. 

Visually, the distribution (\ref{a1}) displays sinusoidal tilt fringes when projected as a real-valued image
$\Re[a(x,y) e^{i \gamma}]$ for a chosen phase offset $\gamma$, such as shown in Fig.~\ref{fig1}(b). 
The fringes are similar to those that 
curvilinear rulings often produce as moir\'{e} patterns, such as the examples shown in Figs.~1(c)--(e). 
The fringes are oriented along the in-plane projection 
$\Delta {\bf k_0} = u_0 \, \hat{x} + v_0 \, \hat{y}$ 
of the radiation wavevector ${\bf k_0}$ with a spacing given by the in-plane wavelength 
$\Delta \lambda_0 = 2 \pi / | \Delta {\bf k_0}| = \pi / (k_0 \sin |\theta_0|).$ 
There are no fringes for broadside steering, and the fringe spacing decreases as the beam steers towards endfire ($\theta_0 = \pm\pi/2$). 

We can synthesize the ideal distribution (\ref{a1}) as a moir\'{e} pattern using the construction (\ref{asm}) as follows. 
To proceed, we will focus on the artificially simple case of a source and a mask that have uniform amplitudes 
\begin{align} 
s_1(x,y) 	&= s_0 \, \exp[{i g_s(x,y)}] \, w(x,y)	\label{s1} \\ 
m_1(x,y) 	&= m_0 \, \exp[{-i g_m(x,y)}]  		\label{m1}
\end{align} 
but phases that vary according to real-valued argument functions $g_s(x,y)$ and $g_m(x,y)$. 
Here, the aperture stop is included with the source for simplicity. 
Using these with (\ref{asm}) and (\ref{a1}), the phase 
arguments must satisfy 
\begin{align} 	\label{textbookresult}
g_s(x,y) - g_m(x,y) = u_0 x + v_0 y + w_0  \pmod{2 \pi}
\end{align} 
within the aperture,  
and the amplitudes must satisfy 
$m_0 s_0 = a_0 \exp({-i w_0})$ 
for the uniform phase offset $w_0$. 
For this case, the steering and its fringes depend only on the arguments and not on the amplitudes. 
Parametrically varying either or both of these arguments then generates dynamic beam steering.

Following Amidror \cite{amidror1}, 
this straightforward result is equivalent to the general case of synthesizing a periodic $(1,-1)$-moir\'{e} from two curvilinear rulings that are repetitive but not necessarily periodic: 
The relationship (\ref{textbookresult}) is identical to the ``fundamental moir\'{e} theorem'' in 
Amidror \cite{amidror1} 
[c.f., Eq.~(10.35), p.~334] 
in which the argument functions 
$g_s(x,y)$ and $g_m(x,y)$ 
serve as bending functions (coordinate transforms) 
that create curvilinear rulings. 
Figs.~1(c)--(e) highlight the nodes of example $(1,-1)$-moir\'{e} patterns from such curvilinear rulings. 
Note that broadside steering requires destructive interference to suppress this $(1,-1)$-moir\'{e} pattern by placing its impulse at the origin ($u_0 = v_0 = 0$). 

The simple case of (\ref{s1}) and (\ref{m1}) using complex-valued functions with single exponential terms (or space harmonics) led to only one interference term and nothing else. 
In practice, the construction (\ref{asm}) generally produces multiple interference terms that can be grouped and indexed as $(n,m)$-moir\'{e} patterns, following the approach of Amidror \cite{amidror1}. 
Typically, only a few of these patterns fall within the visibility window. 
As an example, note that if only the real parts of the source (\ref{s1}) and mask (\ref{m1}) are used, then (\ref{asm}) produces four terms that can be indexed as $(1,-1)$- and $(1,1)$-moir\'{e} patterns. 
The new $(1,1)$-moir\'{e} would be suppressed if it is outside the visibility window.

\section{Antenna examples} 
This section describes six examples of antennas whose beam steering can be interpreted as one of either the three moir\'{e} patterns shown in Figs.~1(c)--(e) or two additional patterns to come below. 
For all but two examples, the main aspects of steering reduce to the simple case presented in the last section, with the steering following from phase arguments. 
{As in the last section, the analysis primarily models the main lobe steering and neglects higher-order detail.} 
Each example demonstrates dynamic control of the steered direction by parametric reconfiguration operations on the source or mask. 

The first and second examples use existing antennas to illustrate steering controlled by scaling operations.  
These antennas are very well understood, so their non-standard treatment here is included both to demonstrate a connection with the moir\'{e} effect and to prepare for the remaining four examples. 

The third and fourth examples illustrate steering controlled by rotating operations. 
One of these is our recent work \cite{stingarray} and the other is a related antenna. 
Both are potential candidates for next-generation satellite constellation user terminals. 
The source and mask distributions of these, as well as the first example, have impulsive nailbed spectra, which allows an interpretation of beam forming and steering with a wavevector diagram for $\Delta {\bf k_0}$. 

The fifth and sixth examples are inspired by known moir\'{e} patterns that have not been extensively explored for use in antennas. 
They illustrate steering controlled by translation or rotation operations. 
Their source and mask distributions are not impulsive nailbeds, but instead outline the special case of identical or strongly correlated mask and source pairs, similar to that of the second example. 
This allows an interpretation of their beam forming and steering as 
a cross correlation, 
like that producing Glass patterns. 

Our focus on linear coordinate-transformation operations like scaling, rotating, and translating follows from their relative ease of mechanical implementation. 
Additionally, they form the largest class of operations (affine transformations) known to have convenient Fourier transform properties, as summarized in Appendix~\ref{AppA}. 
However, many other control operations are possible and could potentially be implemented with reconfigurable antennas or metamaterials. 
For reference, 
Appendix~\ref{AppA} also provides Fourier transforms relevant to the examples to follow.

\begin{figure}[t]
\centering
\includegraphics[width=0.99\columnwidth]{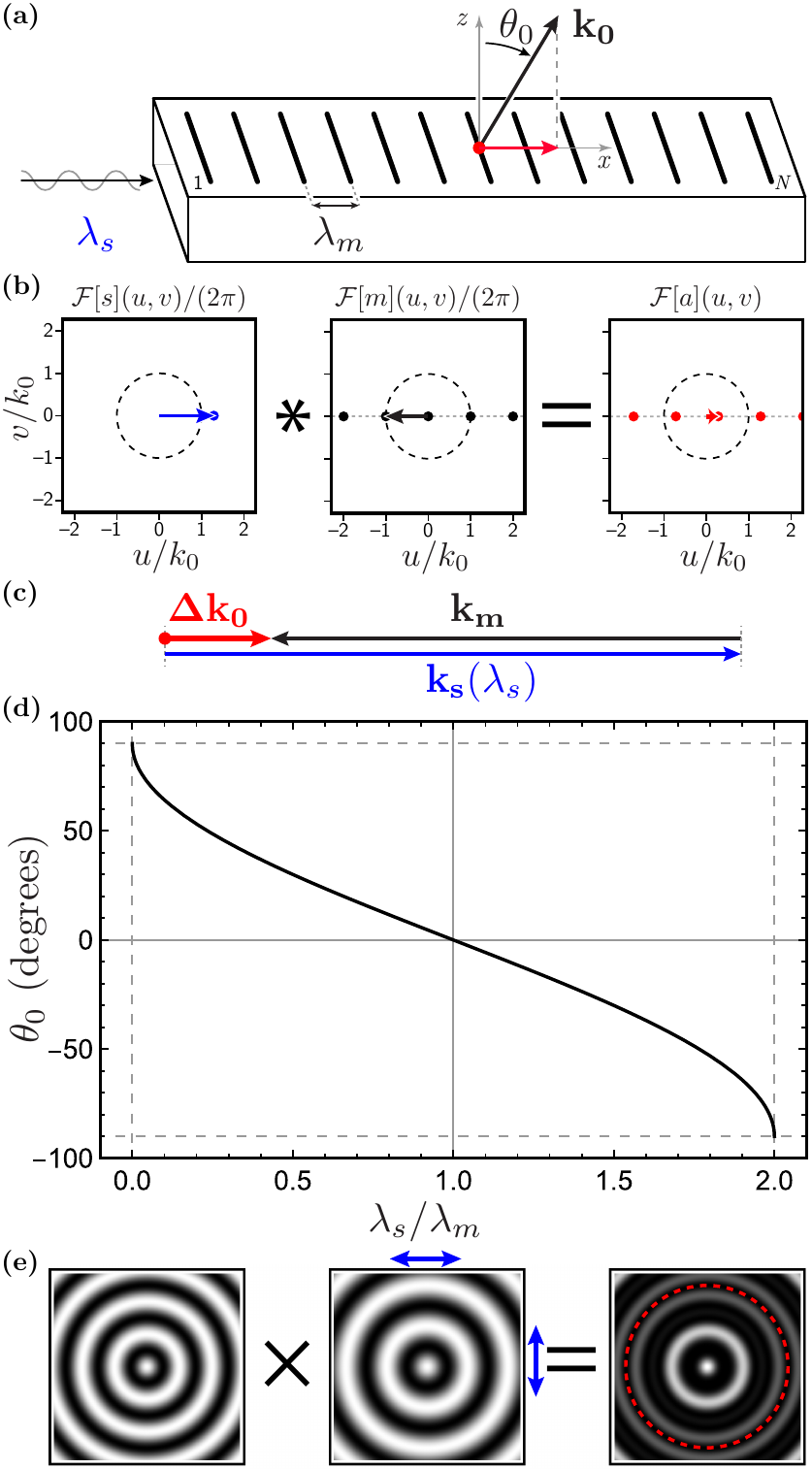}
\caption{ 	\label{fig2} 
Steering by scaling corresponding to Fig.~\ref{fig1}(c). 
(a) Setup of the first example, an array of waveguide slots known as a one-dimensional periodic leaky wave antenna (1D pLWA). 
(b) Sketch of the spectral convolution (\ref{convasm}). 
Dots are impulse locations, circles are visibility windows, and lines are impulse comb axes. 
(c) Vector diagram for the tilt fringe wavevector, using vectors from (b). 
(d) Steering angle $\theta_0$ for the main lobe. 
(e) Moir\'{e} pattern for the related second example (2D bull-eye pLWA), as discussed in the text. 
The red dashed circle highlights a fringe node. 
}
\label{overview}
\end{figure}

\subsection{Steering by scaling} 

This section presents two introductory examples 
of steering by scaling that are related to the moir\'{e} pattern in Fig.~\ref{fig1}(c), which is a moir\'{e} magnification of a scaling \cite{amidror1}. 
Similar patterns occur with identical structures if combined with perspective. 
{These two idealized examples serve to demonstrate a connection with the moir\'{e} effect and to prepare for the remaining examples.} 

\subsubsection*{Example 1} 

For the first example, consider 
transmission from an {idealized} array of waveguide slots as sketched in Fig.~\ref{fig2}. 
While this slot-array antenna is traditionally known as a one-dimensional periodic leaky wave antenna (1D pLWA), it does not require a fast traveling wave like most leaky wave antennas \cite{kraus}. 
It consists of a uniformly spaced 1D array of $N$ slots cut in a waveguide with spacing $\lambda_m = 2 \pi / k_m$. 
It steers the elevation angle $\theta_0 = \theta_0(\lambda_s)$ for fixed azimuth $\phi_0=0$ by controlling the wavelength $\lambda_s = 2 \pi/k_s$ of radiation traveling in the guide about $\lambda_s \approx \lambda_m$. 
Parametrically varying $\lambda_s = \lambda_s(t)$ dynamically steers the beam. 
For convenience, let $k_0 = k_s$. 

We can recover the standard result for its steering considering an array distribution for the slots. 
This distribution can be synthesized with a source 
\begin{align} 	\label{s2}
s_2(x,y) 	&= s_0 \exp({i k_s x}) 
\end{align} 
representing the amplitude of input radiation in the guide at any given moment in time
and a mask 
\begin{align} 	\label{m2}
m_2(x,y) 	&= m_0 \, \delta(y) \sum_{n=0}^{N-1} \delta(x - n \lambda_m)  
\end{align} 
representing the slot array positions. 
Together with (\ref{asm}), the distribution $a_2(x,y) = s_2(x,y) m_2(x,y)$ models the positions and driving excitations of the slots. 
For simplicity, attenuation is neglected in the guide, though we will return to attenuation below. 
Likewise, the slot excitation amplitudes are often carefully tailored, but for simplicity such tapering is neglected here. 
It can be included in the model by suitably modifying the source and mask.

The steering mechanism can be interpreted as moir\'{e} generation like that in Fig.~\ref{fig1}(c) as follows. 
The source is of the form (\ref{s1}) and is a complex-valued version of the sinusoidal rulings in Fig.~\ref{fig1}(c) with a spacing that varies with $\lambda_s$. 
Likewise, the mask is equivalent to a sum of terms of the form (\ref{m1}) that are also sinusoidal rulings. 
The steering is unchanged if we keep only the term 
(impulse) producing the main lobe, 
\begin{align} 		\label{m2approx}
m_2(x,y) \approx (m_0 / \lambda_m) \delta(y) \exp({-i k_m x}) w(x), 
\end{align} 
which has a spacing that is fixed by $\lambda_m$. 
This follows from using a Fourier cosine series to rewrite the mask as a sum of $(N-1)$-th harmonics of the array width. 
Alternatively, this follows from inverse transforming the relevant mask impulse using (\ref{A1fix}). 
Here, 
the aperture stop 
$w(x) = \text{rect}\{x/[(N -1)\lambda_m] - 1/2\}$ 
in terms of the rectangular function $\text{rect}(x)$ that is unity for $x \in [-1/2,1/2]$ 
and zero elsewhere. 

The product (\ref{asm}) of (\ref{s2}) and (\ref{m2approx}) generates an ideal distribution (\ref{a1}) with 
$(u_0,v_0) = (k_s-k_m,0)$. 
The beam angles (\ref{theta0}) and (\ref{phi0}) are then the conventional result 
\begin{align}	\label{1DpLWA}
\theta_0(\lambda_s) 	&= \arcsin [ 1 - (\lambda_s/\lambda_m) ] 
\end{align} 
and $\phi_0 = 0$ 
as shown in Fig.~\ref{fig2}(d) {(c.f., Eq.~16 on p.~760 of \cite{kraus} with $m=-1$)}. 
These angles are equivalent to the large-scale tilt fringes shown in Fig.~\ref{fig1}(c), which are a moir\'{e} magnification of the difference between the source and mask ruling spacings, $\lambda_s$ and $\lambda_m$. 

In the spectral domain, the moir\'{e} generation follows from the convolution (\ref{convasm}) of impulsive spectra as sketched in Fig.~\ref{fig2}(b). 
The source has a single delta-function impulse at 
$(u,v) = (k_s,0)$. 
The mask (\ref{m2}) is a truncated Dirac comb (or Shah function) with spacing $\lambda_m$, and its spectrum is a Dirac comb of spacing $k_m$ broadened by convolution with the aperture-stop spectrum.
The convolution (\ref{convasm}) offsets the mask impulse positions by subtracting that of the source impulse, following the Fourier shift theorem. 
The location of the impulse nearest the origin then follows from the wavevector diagram 
$\Delta {\bf k_0} = {\bf k_s} - {\bf k_m}$
of Fig.~\ref{fig2}(c), 
and is positioned at $\Delta {\bf k}_0$ in k-space as shown in Fig.~\ref{fig2}(b). 
The two neighboring terms in the mask expansion are repositioned near the rim of the visibility window, and produce additional lobes when $\lambda_s \neq \lambda_m$ that can also be interpreted as moir\'{e} patterns. 
These lobes result from the slot array spacings being insufficiently dense. 

In a real transmitting antenna, the source amplitude would decrease along the guide as it loses energy to radiation. 
For modest attenuation, 
this does not change the steering significantly for the same reason that varying the intensity across one of the input images in Fig.~\ref{fig1}(c) would not significantly alter the large-scale fringes in the output image. 
Quantitatively, multiplying the source by a decaying exponential to model attenuation and using (\ref{attenuation}) leads to a convolution of the source spectrum with a sharp Lorentzian, which has little effect on impulse positions in the final spectra like that of the main lobe. 
However, this can affect beam width and side lobe levels just like tapering, which can be modeled similarly. 
For this reason, we do not consider attenuation below, though it must be accounted for to select the aperture size and to characterize the full performance of a real antenna.

\subsubsection*{Example 2} 
For the second example, consider scaling in two dimensions. 
The same $\theta_0$ steering results if we bend the linear rulings of the previous example to form radial rulings as follows. 
Consider transmission from an {idealized} array of concentric circular slots in a waveguide with uniform radial spacing centered about a waveguide feed. 
Such an antenna produces a conical beam peaked at $\theta_0$ across all $\phi_0$ \cite{2DpLWA}. 
Parametrically varying $\lambda_s = \lambda_s(t)$ then dynamically steers this conical beam. 
While this is known as a two-dimensional (2D) annular pLWA or ``bull-eye'' antenna, it does not require a fast traveling wave. 
Therefore, for convenience, let $k_0 = k_s$. 

We can recover its steering by modeling the guide with the source 
\begin{align}	\label{s3}
s_3(x,y) = s_0 \exp\left( i k_s \sqrt{x^2 + y^2}\right)  
\end{align} 
and the slots with the mask 
\begin{align}	\label{m3}
m_3(x,y) = m_0 \sum_{n=1}^N  \delta\left(\sqrt{x^2 + y^2} - n \lambda_m\right) 
\end{align} 
which has a spacing of $\lambda_m$. 



In the spectral domain, the moir\'{e} generation follows from the convolution of impulsive but not nailbed spectra. 
The source is an impulsive circle of radius $k_s$ at the boundary of the visibility window. 
The mask is a series of concentric impulsive circles about an impulse at the origin, with radii near integer multiples of $k_m$, broadened by the aperture stop. 
In the convolution with the source, the smallest of these circles produces an impulsive ring of radius $\sqrt{u_0^2 + v_0^2} \approx |k_m - k_s|$. 
Just as for the 1D pLWA, the steering is unchanged if we keep only a term of the form 
\begin{align}	\label{m3approx}
m_3(x,y) \propto \exp\left(- i k_m \sqrt{x^2 + y^2}\right) w(x,y) 
\end{align} 
where 
the aperture stop is circular, $w(x,y) = \text{circ} [ \sqrt{x^2+ y^2}/(N \lambda_m) ]$, 
in terms of $\text{circ}(x) = \text{rect}(x+1/2)$. 
The form (\ref{m3approx}) is a numerical approximation that produces a similar, though not identical, impulsive circle as the smallest mask circle 
producing the main lobe, so captures the steering but not fine details. 
Note that the mask impulse near the origin may produce additional lobes during steering from its convolution with the source impulsive circle. 

The product of (\ref{s3}) and (\ref{m3approx}) does not generate an ideal distribution (\ref{a1}), except for the broadside case of $\lambda_m=\lambda_s$. 
Instead, it generates a new circular ruling that leads to a conical beam about broadside. 
This new ruling can be interpreted as the moir\'{e} generation shown in Fig.~\ref{fig2}(e). 
In the spectral domain, the source and simplified mask spectra are each an impulsive ring about the origin with radius $\sqrt{u_0^2 + v_0^2} = k_s$ or $k_m$, respectively. 
Their convolution (\ref{convasm}) is equivalent to a cross correlation of 
$\mathcal{F}[s_3](u,v)$ with 
$\mathcal{F}[m_3]^*(-u,-v).$ 
This cross correlation produces an impulsive ring with radius $|k_s - k_m|$, leading to a cone beam with the same peak elevation angle (\ref{1DpLWA}) as in the last example, agreeing with Comite {\it et al.} \cite{2DpLWA} {(c.f., page 3, third paragraph, noting $\beta_{-1} = k_0 - 2 \pi/\lambda_m$ for $\alpha = 0$)}. 
In practice, for a transmitting antenna, the source amplitude would decrease radially from both attenuation and conservation of energy during propagation in the guide. This is ignored here and subsequently as discussed above, but again must be accounted for to capture the full performance of a real antenna. 


\subsection{Steering by rotating} 
This section presents two examples of steering by rotating related to the moir\'{e} pattern in Fig.~\ref{fig1}(d), which is a moir\'{e} magnification of a rotation \cite{amidror1}. 
The first is a prototype mechanically steered array (MSA) demonstrated in our recent work \cite{stingarray}. 
The second is a variable inclination continuous transverse stub (VICTS) antenna originally proposed in Milroy \cite{VICTS1}, developed in \cite{victs-1,victs-2,victs-3,victs-4,victs-5,victs-6}, and also modeled in \cite{stingarray}. 
Both of these are potential candidates for low-cost and low-power satellite constellation user terminals. 
The analysis extends to some Risley and related antennas such as in \cite{2020:baba,2017:afzal,2013:gagnon}.

\begin{figure}[t]
\centering
\includegraphics[width=0.99\columnwidth]{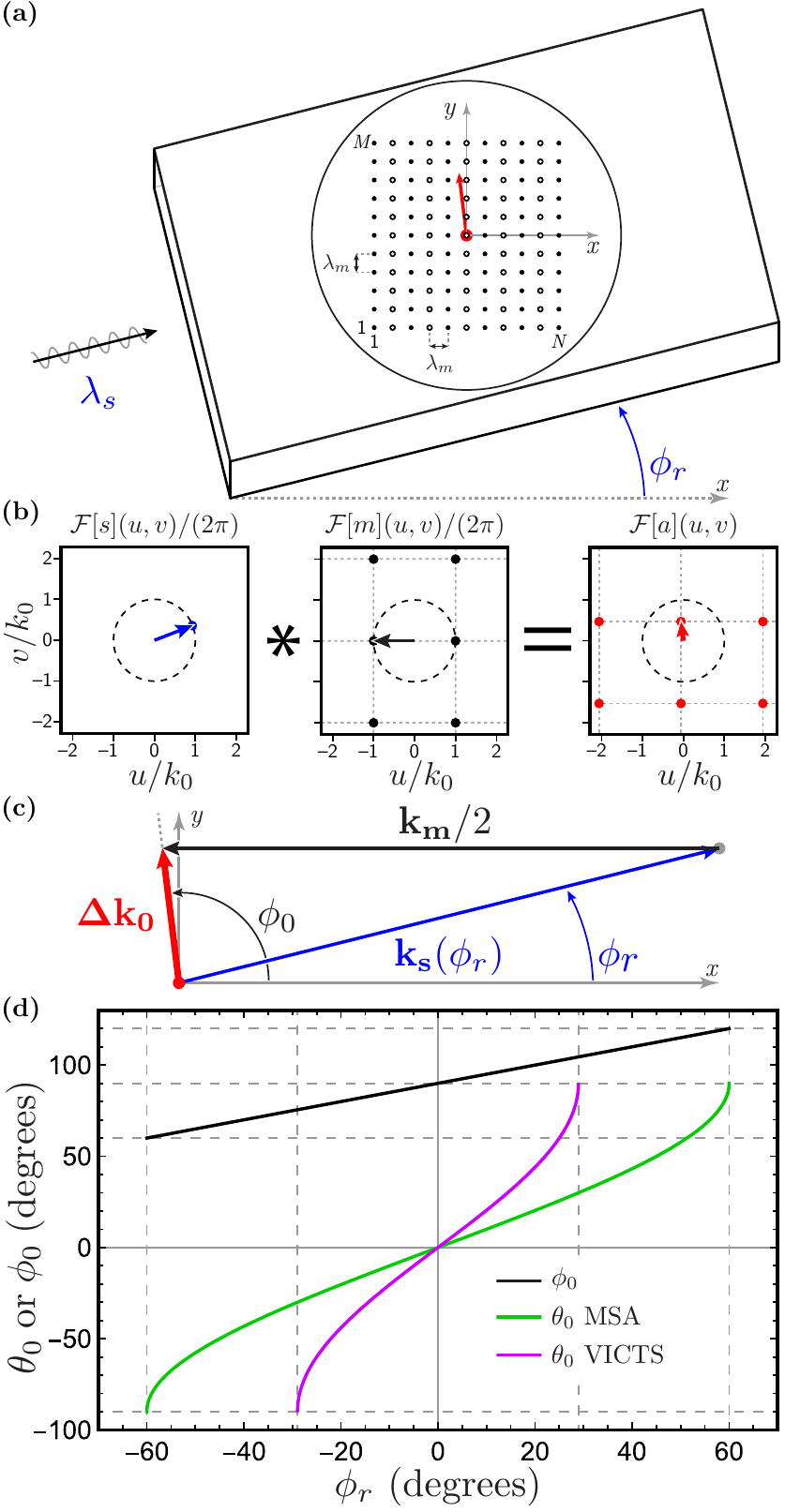}
\caption{ 	\label{fig4} 
Steering by rotating corresponding to Fig.~\ref{fig1}(d). 
(a) Setup of the mechanically steered antenna (MSA) \cite{stingarray} of the third example. 
The VICTS antenna of the fourth example is similar, as described in the text. 
(b) Sketch of the spectral convolution (\ref{convasm}). 
Dots are impulse locations, circles are visibility windows, and lines are impulse comb axes. 
(c) Vector diagram for the tilt fringe wavevector, using vectors from (b). 
(d) Steering angles for the main lobes of the example MSA and VICTS antennas. 
}
\label{overview}
\end{figure}

\subsubsection*{Example 3} 

First, consider the prototype MSA sketched in Fig.~\ref{fig4}(a). 
This antenna consists of a phased array aperture with elements that sample the fields beneath their locations in a waveguide. 
The waveguide provides transverse illumination at a mechanically variable angle. 
The MSA steers the elevation angle $\theta_0 = \theta_0(\phi_r)$ by controlling the relative angle $\phi_r$ between the guide and array, which also induces an azimuth shift $\phi_0 = \phi_0(\phi_r)$. 
Parametrically varying $\phi_r = \phi_r(t)$ dynamically steers the beam. 
For full azimuth control, both the guide and array are rotated, which is not considered here. 
For convenience, let $k_0 = k_s$. 

We can recover the steering of the MSA \cite{stingarray} considering the array distribution, which follows closely the treatment of the first example. 
This distribution can be synthesized with a source 
\begin{align} 	\label{s4}
s_4(x,y) 	&= s_0 \exp \{ i k_s [ \cos(\phi_r) x + \sin(\phi_r) y ] \} 
\end{align} 
representing the guide radiation and a mask 
\begin{align} 	\label{m4}
m_4(x,y) 	&= m_0 \sum_{n=0}^{N-1} \sum_{m=0}^{M-1} (-1)^n \delta(x - n \lambda_m) \delta(y - m \lambda_m) 
\end{align} 
representing the element positions and fixed $(-1)^n$ phase offsets from in-plane element rotations \cite{stingarray}. 
For simplicity, 
attenuation is neglected in the guide and the elements follow a uniform $N \times M$ rectangular layout of spacing $\lambda_m = 2 \pi / k_m$. 
Likewise, element excitation amplitudes are often carefully tailored in phased arrays \cite{victs-6}, but for simplicity such tapering is neglected here. It can be included by suitably modifying the source and mask.

The steering mechanism can be interpreted as moir\'{e} generation like that in Fig.~\ref{fig1}(d) as follows. 
Just as with (\ref{m2}), the mask is equivalent to a sum of terms of the form (\ref{m1}) that are sinusoidal rulings. 
The term (impulse) producing the main lobe is 
\begin{align} 	\label{m4approx}
m_4(x,y) 	&\propto \exp(-i k_m x / 2) w(x,y) 
\end{align} 
which has a fringe spacing fixed at $2 \lambda_m$ instead of $\lambda_m$ from the intentional $(-1)^n$ phase control. 
Here, the aperture stop $w(x,y) = \text{rect}\{x/[(N-1)\lambda_m] - 1/2\}\text{rect}\{y/[(M-1)\lambda_m] - 1/2\}$.  
The source is of the form (\ref{s1}) and is a sinusoidal ruling with a spacing fixed by $\lambda_s$ but a direction that varies with $\phi_r$. 

The product of the source (\ref{s4}) and mask (\ref{m4approx}) generates an ideal distribution with 
$(u_0, v_0) = (k_s \cos[\phi_r] - k_m/2, k_s \sin[\phi_r])$.  
Noting that there is a 180$^\circ$ rotational symmetry, we may consider $\phi_r \in [-\pi/2,\pi/2]$. 
For the case that $k_0 = k_s = k_m/2$, a main lobe exists when $\phi_r \in [-\pi/3,\pi/3]$ 
with beam angles 
\begin{align} 
\theta_0(\phi_r) 	&= \arcsin [ 2 \sin(\phi_r/2) ] \\ 
\phi_0(\phi_r) 		&= (\phi_r + \pi)/2  \label{MSAphi}
\end{align} 
after removing a discontinuity about broadside. 
This agrees with the {results} in our recent work \cite{stingarray} {(c.f., Eqs.~12-13)}. 
At the extremes, $\theta_0(\pm \pi/3) = \pi/2$ and the next terms in an expansion for the mask begin to produce lobes, which can also be interpreted as moir\'{e} patterns. 

In the spectral domain, the moir\'{e} generation follows again from the convolution of impulsive spectra as sketched in Fig.~\ref{fig4}(b).  
This leads to the wavevector diagram $\Delta {\bf k_0} = {\bf k_s} - {\bf k_m}/2$ in Fig.~\ref{fig4}(c) that resembles the Laue diffraction condition for a scattering vector $\Delta {\bf k} = \Delta {\bf k}_0 - {\bf k_s}$ to equal a reciprocal lattice vector $- {\bf k_m}/2$ of the mask \cite{kittel}. 
The source spectrum has a single impulse at $(u,v) = (k_s,0)$ as before. 
The mask (\ref{m4}) is a truncated combination of Dirac combs. 
The mask spectrum is a product of impulse combs with spacings of $k_m$ along both $\hat{u}$ and $\hat{v}$, broadened by convolution with the aperture-stop spectrum. 
However, the phase control factor $(-1)^n$ shifts the impulse locations along $\hat{u}$ by $k_m/2$, as shown in Fig.~\ref{fig4}(b). 

\subsubsection*{Example 4} 
Another example of steering by rotation is the VICTS antenna,  
which resembles the MSA in Fig.~\ref{fig4}(a) but has slots instead of elements. 
This antenna consists of a 1D array of long, parallel slots that couple to the fields beneath their lengths in a waveguide. 
Just as with the MSA, the waveguide provides transverse illumination at a variable angle $\theta_r$, which steers the elevation angle and also induces an azimuth shift. 
Parametrically varying this angle dynamically steers the beam. 
For full azimuth control, both the guide and array are rotated. 

We can model the VICTS steering considering the array distribution 
from a source modified from (\ref{s4}) to include a factor of 2, 
\begin{align} 	\label{s5}
s_5(x,y) 	&= s_0 \exp \{ i 2 k_s [ \cos(\phi_r) x + \sin(\phi_r) y ] \},  
\end{align} 
to represent radiation in a dielectric slow-wave guide with wavenumber $2 k_s = 2 k_0$ and a mask representing the slot positions 
\begin{align} 	\label{m5}
m_5(x,y) 	&= m_0 \, \delta(y) \sum_{n=0}^{N-1} \delta(x - n \lambda_m).  
\end{align}  
As before, both the source and mask can be interpreted as sinusoidal gratings. 
The term (impulse) in the mask 
producing the main lobe 
remains of the form (\ref{m4approx}) but with $k_m$ instead of $k_m/2$. 


The product of the source (\ref{s5}) and mask (\ref{m4approx}) generates an ideal distribution with 
$(u_0,v_0) =  (2 k_s \cos[\phi_r] - k_m, 2 k_s \sin[\phi_r]).$ 
Noting that there is a 180$^\circ$ rotational symmetry, we may consider $\phi_r \in [-\pi/2, \pi/2]$. 
For the case that $k_0 = k_s = k_m / 2$, a main lobe exists when $|\phi_r| \leq \arccos(7/8) \approx 29^\circ$ 
with elevation angle 
\begin{align}
\theta_0(\phi_r) &= \text{sgn}(\phi_r) \arcsin \left( 2 \sqrt{2[1 - \cos (\phi_r)]} \right) 
\end{align} 
and the same azimuth angle as (\ref{MSAphi}), after removing a discontinuity about broadside. 
At the extremes, $\theta_0[\pm \arccos(7/8)] = \pi/2$ and the next terms in the expansion for the mask begin to produce additional lobes. 
This agrees with the analysis in our recent work \cite{stingarray} {(c.f., Eqs.~14-15 with $p=1$) and \cite{victs-4} (c.f., Eqs.~12-13, accounting for rotation convention and k-space units)}. 

In the spectral domain, the moir\'{e} generation follows again from the convolution of impulsive spectra, leading to a similar wavevector diagram as that in Fig.3(b), but with 
$\Delta {\bf k_0} = 2 {\bf k_s} - {\bf k_m}$. 

\subsubsection*{Discussion} 

Before we continue, let us discuss the last two examples and how the MSA of our recent work \cite{stingarray} illustrates using a moir\'{e} pattern to generate an antenna design. 
As outlined above, the MSA steers using the same type of moir\'{e} pattern that is shown in Fig.~\ref{fig1}(d) as the VICTS as well as some Risley and related antennas \cite{2020:baba,2017:afzal,2013:gagnon}. 
That is, while these antennas appear different because of their implementations, the connection with the moir\'{e} effect highlights that 
their steering mechanisms are physically nearly identical. 
This type of pattern was chosen as a starting point to synthesize the MSA because of its suitability for compact, flat panel antennas with wide-angle beam steering. 

A key design driver of the MSA was to reduce cost \cite{stingarray}. 
For the VICTS, the choice to use slots requires an expensive slow-wave source guide. 
This is because the slots must be placed sufficiently densely ($\lambda_0/2$ spacing) to avoid unwanted lobes, such as occur in the first example, and as implemented there is no convenient method for slot-specific phase control in the mask. 
Therefore, the source must have a matching spatial wavelength (sinusoidal ruling spacing) near $\lambda_0/2 < \lambda_0$. 

To reduce cost, the MSA uses a less expensive hollow source guide. 
This choice constrains the source spatial wavelength to $\lambda_0$ and thus the source to (\ref{s4}). 
To generate the desired pattern, the mask must have a corresponding term (\ref{m4approx}) with a matching spatial wavelength near $\lambda_0$ along one direction. 
The MSA chose to implement this using a dense ($\lambda_0/2$ spacing) array of elements sampling the source beneath their locations both to avoid unwanted lobes and because there is significant freedom to tailor the phases (and amplitudes) of such elements. 
Specifically, the MSA prototypes in \cite{stingarray} used in-plane element rotations to implement the mask (\ref{m4}), which are convenient for circularly polarized radiation. 
These rotations allow the term (\ref{m4approx}) to have a longer wavelength than the array spacing. 
Note that such an approach could synthesize more complex masks with nearly full control of the mask spectrum within the visibility window. 

The connection with the moir\'{e} effect readily exposes that the combination of a fixed array spacing but variable guide wavelength introduces additional dispersion and squint issues in MSA and VICTS antennas \cite{stingarray}. 
In particular, for the case of $k_s$ close but not equal to $k_m / 2$, 
there is a keyhole blocking transmission near broadside even for static beam steering, which does not seem to be appreciated in the literature. 
This occurs because destructive interference is not able to completely suppress the moir\'{e} pattern producing the main lobe. 
This keyhole arises if, for example, the guide wavelength (frequency) is varied away from matching. 
This is important in practice if dissimilar wavelengths are used, say for transmit and receive, and could potentially be mitigated by combining separate arrays of different elements on the same mask. 
%
%

\subsection{Steering by translating} 
This section presents two  
beam-steering mechanisms connected to well-known moir\'{e} patterns that are dynamically controlled by translation. 
The first is equivalent to the moir\'{e} pattern highlighted in Fig.~\ref{fig1}(e), and the second to a related pattern to come below. 
Both mechanisms form and steer a main lobe by correlation, similar to the second example.

\subsubsection*{Example 5} 
The pattern in Fig.~\ref{fig1}(e) is a traditional example of ``Schuster fringes'' \cite{schuster,amidror1} generated from translation. 
This particular pattern comes from multiplying two complementary circular zone rulings.  
However, this pattern belongs to a broad class of similar fringes with 1D or 2D control that arise from multiplying elliptical, hyperbolic, parabolic, or spiral zone rulings \cite{amidror1,spiral}. 
Such zone rulings can be synthesized and controlled themselves as moir\'{e} patterns \cite{amidror1,patorski}. 

\begin{figure}[t]
\centering
\includegraphics[width=0.99\columnwidth]{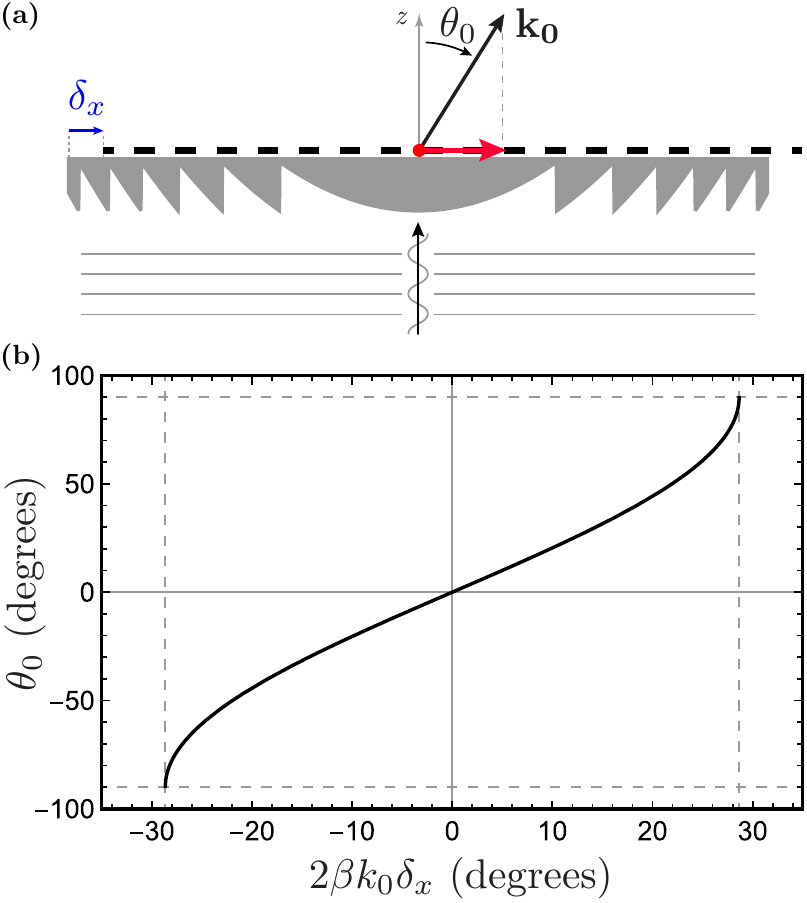}
\caption{ 	\label{fig5} 
Steering by translation with zone patterns corresponding to Fig.~\ref{fig1}(e) for the fifth example. 
(a) Cross section of a potential implementation using normal plane-wave illumination of a translating phased-array board (dashed line) after a lens (grey). 
(b) Steering angles for the main lobe. 
}
\label{overview}
\end{figure}

For the fifth example, consider a source that is a complex-valued version of a sinusoidal zone ruling, 
\begin{align} 
s_6(x,y) 	&= s_0 \exp \left[i \beta k_0^2 \left(x^2 + y^2\right) \right] w(x,y) 
\end{align} 
where the ruling parameter $\beta$ controls how the phase grows quadratically with the radius $\sqrt{x^2 + y^2}$. 
Consider a mask with a complementary ruling 
\begin{align} 
m_6(x,y) 	&= m_0 \exp \left\{ - i \beta k_0^2 \left[(x - \delta_x)^2 + (y-\delta_y)^2\right] \right\} 
\end{align} 
that has a larger extent so its center can be offset from the source by the control parameters $\delta_x$ and $\delta_y$. 
Their product generates an ideal distribution with 
$(u_0, v_0) = (2 \beta k_0^2 \delta_x, 2 \beta k_0^2 \delta_y)$, 
because the difference of two identical but offset parabolas is a line. 
In the spectral domain, this is equivalent to a cross correlation of the spectra $\mathcal{F}[s_6](u,v)$ and $\mathcal{F}[m_6]^*(-u,-v)$. 
For $\delta_x=\delta_y=0$, this generates an impulse at the origin. 
Otherwise, translation produces a linear phase gradient in the image that translates this impulse in the spectrum away from the origin following the Fourier shift theorem. 
The beam angles are 
\begin{align} 
\theta_0(\delta_x,\delta_y) 	
	&= \text{sgn}(\beta \delta_x) \arcsin\left( 2\left| \beta \right| k_0 \sqrt{ \delta_x^2 + \delta_y^2 } \, \right)  \\ 
\phi_0(\delta_x,\delta_y) 		
	&= \arctan (\delta_y / \delta_x )
\end{align} 
as shown in Fig.~\ref{fig5} for the case of $\delta_y = \phi_0 = 0$. 
Note that rotations could also be used for steering, such as in the related optical beam-steering mechanism in Bawart {\it et al.} \cite{2020:bawart}.

\begin{figure}[t]
\centering
\includegraphics[width=0.99\columnwidth]{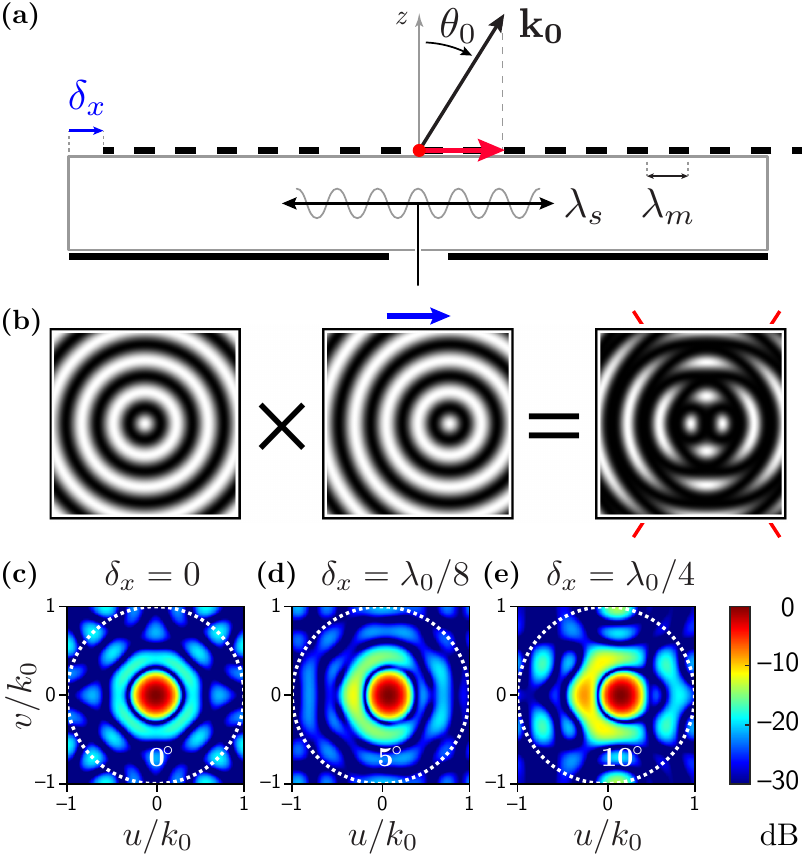}
\caption{ 	\label{fig6} 
Fine steering by translation with circular patterns for the sixth example. 
(a) Cross section of a potential implementation using transverse cavity illumination of a translating phased-array board (dashed line). 
(b) Corresponding moir\'{e} pattern. Red lines highlight fringe nodes. 
(c-e) Squared magnitude of the array factor for an example case described in the text for different offsets $\delta_x$ showing approximate angles $\theta_0$. 
The white dashed circles are visibility windows. 
}
\label{overview}
\end{figure}

The antenna of Lima {\it et al.} \cite{2015:lima} as well as related beam-switched antennas that use translation or rotation of a mask held at a distance from a horn source 
appear closest to an implementation of this steering example. 
This is because the intersection of spherical waves with distant planes approximates circular zone patterns. 
However, the broad class of such fringes include much more than traditional circular-zone patterns, as listed above. 
Therefore, this class of moir\'{e} patterns appears to have not been extensively explored despite its potential application in wide-angle steering. 
In particular, steering by translation of only a mask may reduce the switching times between moving terminals compared to steering by separate rotations of a mask and a source 
as used in the VICTS and MSA. 
{In addition, there is no dynamic keyhole during tracking near broadside (i.e., no rapid rotations to step $\phi_0$ by $\sim\pm\pi$).} 

Even for traditional circular-zone patterns, there are several potential implementations left to investigate. 
In particular, a cavity-backed source similar to Kodnoeih {\it et al.} \cite{2020:kodnoeih} may be attractive for a compact antenna. 
Alternatively, Fig.~\ref{fig5}(a) sketches a straightforward implementation using normal-incident illumination of a near-field dielectric phase-shifting structure similar to a Fresnel lens to synthesize the source as an illumination sampled by a phased array. 
The array could use fixed phase offsets similar to the MSA described above to synthesize the mask. 
Additionally, two such dielectric structures, a transmitarray, or a reflectarray may be possible. 
Or, the zone gratings could be synthesized as moir\'{e} patterns. 

\subsubsection*{Example 6} 
As a sixth example, let us consider a related case that is potentially easier to implement but provides a more limited steering range. 
While this example is not a candidate for applications similar to those of the MSA or VICTS, it is a potential candidate for low-power and low-cost applications requiring fine steering. 
Note that it is challenging to implement a circular zone source with a cavity \cite{2020:kodnoeih}. 
However, a center-fed cavity with radial propagation can approximate a roughly similar circular phase ruling, as in the second example discussed above and as sketched in Fig.~\ref{fig6}(a). 
Such a source can be modeled as 
\begin{align} 
s_7(x,y) 	&= s_0 \exp \left( i k_s \sqrt{x^2 + y^2} \right) w(x,y),  
\end{align} 
which is the same as $s_3(x,y)$ except for including $w(x,y)$. 
A suitable mask of the ideal form (\ref{m1}) could be constructed for any initial direction $(\theta_0, \phi_0)$. 
Translating (or rotating) such a mask would produce fine steering about this initial direction. 
Without loss of generality, consider steering about broadside $\theta_0=0$, for which this mask is 
\begin{align} 
m_7(x,y) 	&= m_0 \exp \left( - i k_s \sqrt{(x - \delta_x)^2 + (y-\delta_y)^2} \right), 
\end{align} 
where $\delta_x$ and $\delta_y$ are translations of the mask with respect to the source. 
For simplicity, let $\delta_y = 0$ and $k_s = k_0$. 

Using these, the construction (\ref{asm}) does not produce the ideal form after translation. 
Instead, for small translations $\delta_x$, it produces a moir\'{e} pattern like that shown in Fig.~\ref{fig6}(b) with curved fringes resembling a hyperbolic grating. 
An approximation for this distribution is 
$a_7(x,y) 	\approx s_0 m_0 \exp \left( { i k_0 \delta_x (x - \delta_x/2)}/ {\sqrt{(x - \delta_x/2)^2 + y^2 }}\right) w(x,y)$. 
The resulting steering roughly scales as $\theta_0 \propto \delta_x/D$ for a circular aperture of diameter $D$. 
However, steering rapidly deteriorates beam quality, with a limited range that roughly scales as $\text{max}|\theta_0| \propto \lambda_0/D$.

Figs.~5(c-e) shows a specific case of a circular source with diameter $D = 4 \lambda_0$ and an oversided, rectangular-grid phased-array mask with element spacings $\lambda_m = \lambda_0/2$ and fixed element phase offsets. 
Here, the plots show the squared magnitude of the array factor normalized to its peak value for different translations $\delta_x$ and resulting steering $\theta_0$ up to roughly $10^\circ$ as labeled. 
As the progression shows, translation introduces significant side lobes, which might be suppressed with amplitude tailoring. 

In the spectral domain, this steering works as follows. 
For no translation, the mask and source have complementary circular impulsive spectra. 
Their convolution creates an impulse at the origin broadened by the aperture-stop spectrum. 
Translation applies a linear phase gradient across the mask spectra, from the Fourier shift theorem. 
If the source were truly an impulsive circle, then this would have little effect. 
However, a finite aperture broadens the impulsive circle of the source, leading to a broadened impulse near the origin after convolution. 
The translation-induced phasing of the mask then distorts the shape of this broadened impulse. 
Thus, a larger steering range is possible for smaller apertures. 

The steering range might be extended along one translation axis with numerical optimization of the mask. 
Note that if the mask is synthesized for an initial direction away from broadside, then rotation can cover all azimuth angles but elevation steering would still have a limited range.

\section{Conclusion} 

In summary, there is an explicit connection between the moir\'{e} effect, beam steering, and antennas that is known in optics and also applies at radio, microwave, and mm-wave frequencies for antennas that use spatial interferences to form beams. 
We show that this connection provides an alternate interpretation of four existing antennas: 
a linear array of waveguide slots (1D pLWA), 
a concentric array of circular waveguide slots (2D annular pLWA), 
the MSA of our recent work \cite{stingarray}, 
and the VICTS. 
Finally, we illustrate how this connection enables the generation and exploration of antenna designs from moir\'{e} patterns 
through the MSA of our recent work as well as through two additional examples. 
We hope that this connection and the perspective it provides will 
be of use to antenna engineers and scientists.

There are many potential avenues future work could explore. 
In the last section, we suggest that there is a potential opportunity to develop compact, low-cost, and fast-switching user terminal antennas for next-generation communication systems based on Schuster fringes and related patterns. 
More broadly, note that while 
the approach here focused on the simple case of planar antennas and the multiplication of two inputs, the moir\'{e} effect also occurs on non-planar surfaces, with different means of nonlinear combination, and with more inputs.  
More complicated source distributions (e.g., higher-order cavity modes) or non-affine control operations (e.g., with metasurfaces) could also be considered. 
Applications with multiple beams, with discrete instead of continuous steering, with different beam shapes, 
or with control besides steering (e.g., focusing) are possible. 
Small-scale interferences may potentially be able to influence superdirectivity. 
Finally, potential applications exist outside optics and electromagnetics, such as acoustics.

\begin{acknowledgments}
This work was supported by Facebook Connectivity, and was 
performed before B.~H.~McGuyer joined Amazon and before Qi~Tang joined Google. 
\end{acknowledgments}

\appendix 

\section{2D Fourier transforms and affine theorem} 
\label{AppA}

For reference, here are Fourier transforms relevant to the examples given above that
assume real-valued quantities $a$ and $b$, $a \neq 0$, and the transform convention (\ref{FT}): 
\begin{align} 
&\mathcal{F}\left[ e^{i(a x + b y)} \right](u,v) 
	= 4 \pi^2 \delta(u-a)\delta(v-b)	\label{A1fix}  \\ 
&\mathcal{F}[\delta(x-a)\delta(y-b)](u,v) 
	= e^{-i(a u + b v)} \\ 
&\mathcal{F}[\text{rect}(a x - b)\delta(y)](u,v) 
	=  \frac{ 2 \text{sgn}(a) e^{\frac{-i b u }{ a}} \sin\left( \frac{ u }{2a} \right) }{ u } \\ 
&\mathcal{F}\left[e^{i a \sqrt{x^2 + y^2}} \, \right](u,v) 
	= \frac{ - 2 \pi }{  a^2 [1 - (u^2 + v^2)/a^2]^{3/2} } \\ 
&\mathcal{F}\left[\text{circ}\left( |a| \sqrt{x^2 + y^2} \right)\right](u,v)  
	= \frac{ 2 \pi J_1 \left( \frac{\sqrt{u^2 + v^2}}{|a|} \right) }{|a| \sqrt{u^2 + v^2}} \\ 
&\mathcal{F}\left[\delta \left( \sqrt{x^2 + y^2} - |a| \right) \right](u,v) 
	= 2 \pi |a| J_0 \left( |a| \sqrt{u^2 + v^2}  \right)  \\ 
&\mathcal{F}\left[e^{- i a (x^2 + y^2)}  \right](u,v) 
	= - i \pi e^{i (u^2 + v^2) / (4 a)} / a.  \\ 
&\mathcal{F}\left[ e^{- |a (x - b)|} \right] (u,v) 
	=  \frac{4 \pi |a| e^{-i b u}}{a^2 + u^2} \, \delta(v) \label{attenuation}
\end{align} 
Here, $\text{rect}(x)$ is unity for $x \in [-1/2,1/2]$ and zero elsewhere, 
$\text{circ}(x) = \text{rect}(x+1/2)$, 
and $J_n(x)$ is a Bessel function of the first kind.


The affine theorem \cite{bracewell:1993,bracewell,amidror1,amidror2} concisely summarizes many properties of Fourier transforms by describing how linear coordinate (affine) transformations are related between the image and spectral domains. 
Consider the linear transformation 
${\bf x'} = {\bf A x} + {\bf x_1}$ 
given explicitly by the coefficients $a, b, c, d, x_1,$ and $y_1$ as 
\begin{align} 
\begin{pmatrix}
x' \\
y' 
\end{pmatrix} 
= 
\begin{pmatrix} 
a & b \\
c & d 
\end{pmatrix} 
\begin{pmatrix}
x \\
y 
\end{pmatrix} 
+ 
\begin{pmatrix}
x_1 \\
y_1 
\end{pmatrix}, 
\end{align} 
which could represent a translation, rotation, reflection, scaling, shear, etc., or a combination of the same. 
Consider also the corresponding transformation 
${\bf u'} = {\bf B u} + {\bf u_1}$ 
that involves the inverse transpose ${\bf B} = {\bf A}^{-T}$ and is given explicitly by the same coefficients with $u_1$ and $v_1$ as 
\begin{align} 
\begin{pmatrix}
u' \\
v' 
\end{pmatrix} 
&= \frac{1}{ad-bc}
\begin{pmatrix} 
d & -c \\
-b & a 
\end{pmatrix} 
\begin{pmatrix}
u \\
v 
\end{pmatrix} 
+ 
\begin{pmatrix}
u_1 \\
v_1 
\end{pmatrix}. 
\end{align} 
Then a general form for the affine theorem is 
\begin{align}
\mathcal{F}[ e^{ - i {\bf u_1} \cdot {\bf x'}} g({\bf x'}) ]({\bf u}) 
	&=  \frac{e^{ i ({\bf x_1} \cdot {\bf u'} - {\bf x_1} \cdot {\bf u_1})} }{ad - bc} \mathcal{F}[g({\bf x})]({\bf u'}), 
\end{align} 
which follows from (\ref{FT}) using the substitution 
$ux + vy	= u' x' + v' y' - {\bf x_1} \cdot {\bf u'}  - {\bf u_1} \cdot {\bf x'}  + {\bf x_1} \cdot {\bf u_1} $
and changing variables of integration.


%

\end{document}